# Optical Frequency Comb Noise Characterization Using Machine Learning

*Giovanni Brajato[1*], Lars Lundberg[2], Victor Torres-Company[2], Darko Zibar[1]*

[1] DTU Fotonik, Technical University of Denmark, Building 343, Ørsteds Pl., 2800 Kgs. Lyngby, Denmark
[2] Photonics Laboratory, Dept. of Microtechnology and Nanoscience, Chalmers University of Technology, SE-41296 Göteborg, Sweden
[*]gibra@fotonik.dtu.dk

**Keywords**: MACHINE-LEARNING, FREQUENCY COMBS, PHASE ESTIMATION

## Abstract

A novel tool, based on Bayesian filtering framework and expectation maximization algorithm, is numerically and experimentally demonstrated for accurate frequency comb noise characterization. The tool is statistically optimum in a mean-square-error-sense, works at wide range of SNRs and offers more accurate noise estimation compared to conventional methods.

## 1. Introduction

Optical frequency combs (OFCs) are envisioned to play a significant role in the next generation of high-speed optical as well as optical-wireless communication systems due to their ability to provide frequency-stable and low phase noise comb lines from a single optical source [1]. The optical phase noise of the individual comb lines is dictated by the optical reference. Importantly, since there is only two degrees of freedom in setting the comb spectrum, the optical phase noise can be correlated among frequency lines. For the application in WDM communication systems [2], accurate noise characterization in terms of phase correlation matrix is essential. The correlation matrix provides information about the phase correlation between comb lines and is important for the design of receiver digital signal processing (DSP) algorithms [3].

In general, it is challenging to obtain the correlation matrix, as accurate optical phase tracking of comb lines is required. Homodyne detection schemes in combination with pulse shaping has been used before [4], but the solution does not measure the correlation matrix in a line-resolved manner, as required in optical communication systems. In a recent work [5], the correlation matrix has been obtained with frequency line resolution based on dual-comb spectroscopy (simultaneous electronic down-conversion with the aid of another frequency comb). This approach (see Fig. 1) captures the amplitude and phase variations of all the comb lines. The signal processing is done based on standard Fourier processing. However, this method is not statistically optimum in a mean-square-error-sense (MSE) and its performance cannot be guaranteed over a wide range of signal-to-noise-ratios (SNRs). Typically, the SNR varies, as the power spectrum density of the frequency comb is not flat. Most importantly, for low-linewidths approaching 100 Hz, the conventional approach for optical phase estimation is inaccurate. It would therefore be useful to have a characterization tool that works over a wide range of SNRs and performs optimum optical phase tracking irrespective of the magnitude of the linewidth.

In this paper, we present a novel machine learning (ML) framework for computation of the phase correlation matrix from dual-comb spectroscopy measurements. The method is based on joint and statistically optimum, in terms of MSE, optical phase tracking. This approach is independent of the magnitude of the linewidth. The framework is investigated numerically and experimentally, and significant advantages over the state-of-the-art method are demonstrated in terms of the accuracy of the estimated correlation matrix and differential phase noise variance.

## 2. Machine learning framework

A detailed schematic of the set-up used of numerical and experimental investigations is shown in Fig. 1. The goal is to perform joint optical phase tracking and extract the correlation matrix of the incoming (source) optical comb that is mixed in a balanced receiver with a strong local oscillator (LO) comb. This approach assumes either that the phase noise properties of the incoming comb and the LO comb are equal, or that the phase-noise contribution of the LO comb can be neglected.

The optical phase tracking is performed after the photocurrent has been sampled by the ADC. Given the sampled photocurrent, $y_k$, statistically optimal phase tracking is obtained by Bayesian filtering [6]. Implementation of the Bayesian filtering framework requires a state-space model that consists of: 1) a measurement equation, which describes the relation between $y_k$ and the time-varying optical phase, $\phi_k$ and, 2) state equation, which describes the evolution of the optical phase difference between the signal and the LO: $\phi_k = \phi_k^{sig} - \phi_k^{LO}$.

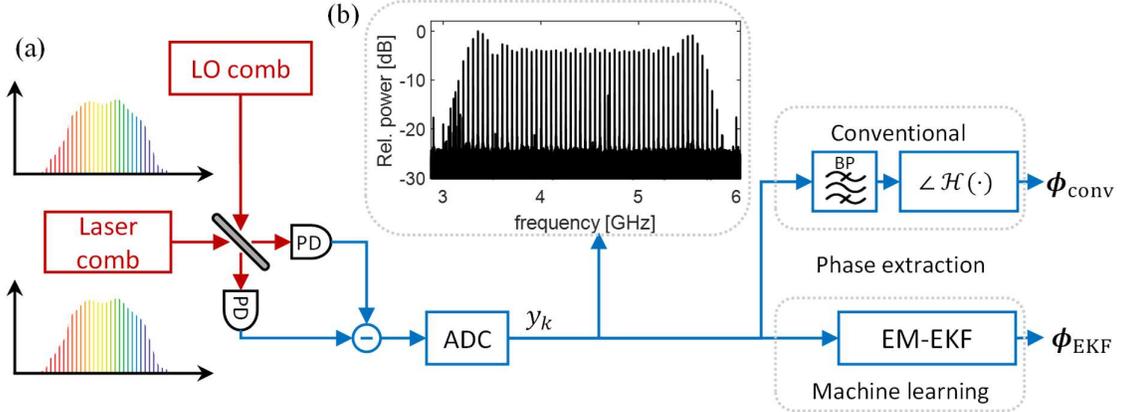

Fig. 1: (a) System set-up for numerical and experimental investigations. Red lines represent optical paths, blue lines represent electrical/digital paths. PD: photodiode. ADC: analog-to-digital converter. BP: band-pass filter. $\angle \mathcal{H}(\cdot)$: phase extractor of the analytic signal computed using the Hilbert transform. EM-EKF: expectation maximization algorithm with extended Kalman filter. $\boldsymbol{\phi}_{\text{conv}}$: phases extracted with conventional method. $\boldsymbol{\phi}_{\text{EKF}}$: phases extracted with machine learning method. (b) Power spectrum density of the downconverted frequency comb in the experiment.

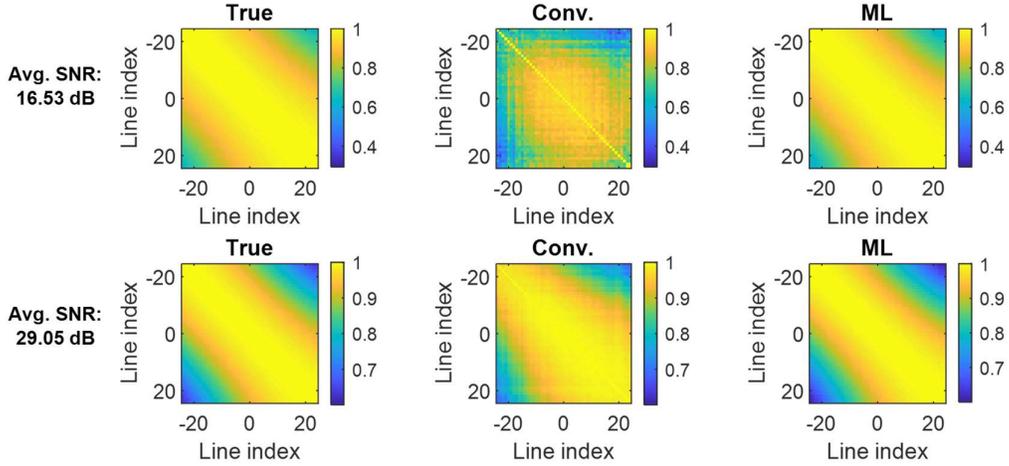

Fig. 2: (Numerical) True correlation matrices (left matrices) shown together with the correlation matrices obtained by the conventional method (central matrices) and the machine learning method (right matrices). Each row is a simulation with different values of the average SNR, indicated on the left.

Our proposed state-space model is the following:

$$\phi_k^1 = \phi_{k-1}^1 + q_{k-1}^1$$
$$\vdots \quad (1)$$
$$\phi_k^M = \phi_{k-1}^M + q_{k-1}^M$$

$$y_k = \sum_m a_m \sin(\Delta\omega_m T_S k + \phi_k^m) + n_k \quad (2)$$

Here, $y_k$ are the discrete-time samples after the ADC, $k$ is an integer representing time, $F_S = 1/T_S$ is the sampling frequency of the ADC and $a_m = 2R\sqrt{P_s^m P_{LO}^m}$ are the line amplitudes. $R$ is the responsivity of the photodiodes, $P_s^m$ and $P_{LO}^m$ are the powers of the signal and LO frequency comb lines. $n_k$ is the measurement noise contribution associated with the shot noise. It is assumed that $n_k$ has Gaussian distribution with zero mean and variance $\sigma_{sn}^2$. The frequency difference between the comb lines is expressed as $\Delta\omega_m$. The finite linewidth of the comb lines can be modelled by assuming the phase noise dynamics (1) as a Markov random walk [7], where $\boldsymbol{q}_k = [q_k^1, \ldots, q_k^M]^\top$ is a Langevin source with covariance matrix $\mathbf{Q}$.

Optical phase tracking can be performed once the state-space model has been defined. The main idea behind Bayesian filtering is to provide a recursive algorithm that computes a statistically optimum joint estimation of phases $\phi_k^m$, for $m = 1 \ldots M$, given $y_k$. The optimum estimates will be the means $\bar{\boldsymbol{\phi}}_k = [\bar{\phi}_k^1, \ldots, \bar{\phi}_k^M]^\top$, where $\top$ represents the transpose operation and $\bar{\boldsymbol{\phi}}_k$ correspond to the phases that minimize the mean squared error (MSE) [8]. In this paper, the Bayesian filtering framework is implemented using the extended Kalman filter (EKF), which we use to compute the mean value $\bar{\boldsymbol{\phi}}_k$. Additionally, the system also includes unknown static parameters that need to be jointly estimated together with the dynamic parameters. The comb relative frequencies $\Delta\omega_m$ and mean amplitudes $a_m$ can easily be extracted by inspection of the signal power spectral density.

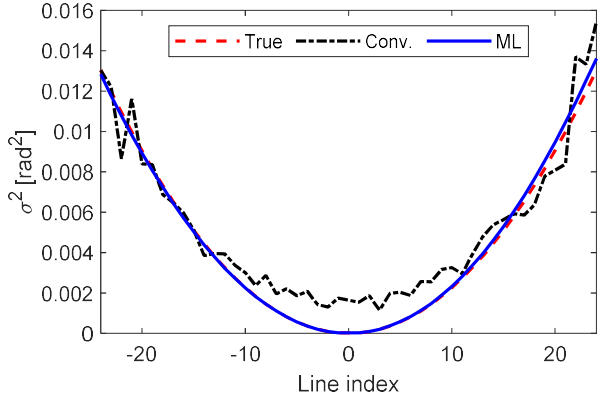

Fig. 3: (Numerical) Empirical variance calculated on the extracted differential phases using machine learning (blue curve) and a conventional method (black curve).

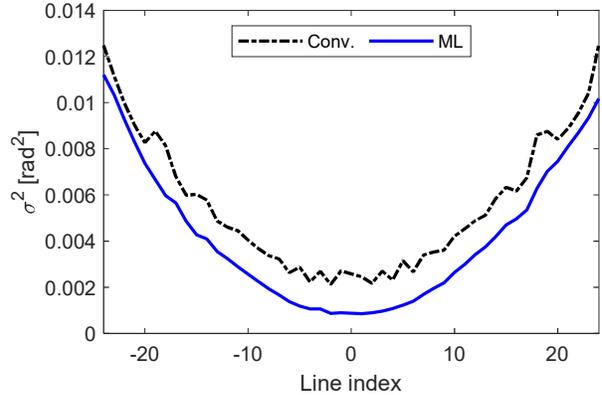

Fig. 4: (Experimental) Empirical variance calculated on the extracted differential phases using machine learning (blue curve) and a conventional method (black curve).

Other parameters such the measurement noise variance $\sigma_{sn}^2$ and the covariance of the phase noise $\mathbf{Q}$ needs to be inferred from the data. To learn them, we decided to use Expectation Maximization (EM) algorithm [9]. The EM iterates over the training data by forward filtering and backward smoothing. At each iteration, the optimal parameters ($\sigma_{sn}^{2(opt.)}$, $\mathbf{Q}^{(opt.)}$) which maximizes the model likelihood on the observed data are returned. The iterative process repeats until convergence.

## 3. Numerical results

The numerical simulation of the system is done by introducing a linear relationship on the phase noise. The generation of a frequency comb with a specific intra-line correlation is done in order to test if the algorithms can recover the same correlation after corrupting the data with white noise samples. Following the same signal structure of electro-optical frequency combs, we generate line-dependent phase noise, i.e. $\phi_k^m = \phi_k^C + m\phi_k^{RF}$. Here, $\phi_k^C$ and $\phi_k^{RF}$ are the carrier and RF phase noise, generated as Wiener processes. The integer line index $m$ takes values in $\{-24, 24\}$, giving 49 lines in total. After generating the signal, we compare the ML method with a conventional technique for phase extraction. It consists of bandpass filters for each comb line, using 30 MHz of bandwidth. For each filtered line, a Hilbert transform is performed to extract the phase of the individual line.

In Fig. 2, we show the true covariance matrix together with the correlation matrix obtained by the conventional and ML method. The average SNR is varied from 16.53 dB to 29.05 dB. The proposed ML framework is able to extract the correlation matrix that is in an excellent agreement with the true one. This is due to its ability to filter out additive measurement noise, a property that is not part of the conventional method. The filtering capabilities of the latter are limited, as it cannot remove the noise within its bandwidth. We can see indeed that the conventional method fails to capture the phase correlation, especially on lower SNR values. Next, we check the differential phase variance extracted using the proposed ML framework with the conventional approach. The differential phase is calculated from the extracted phases as $\Delta\phi_k^m = \phi_k^m - \phi_k^0$. From the way the data is generated, the differential phase true variance is a parabolic function of the line index. In Fig. 3 we can see that our ML method to extract the variance is more accurate than a conventional one, as the ML curve overlaps the true variance curve. The conventional method suffer from measurement noise that affects the variance estimation. Our ML algorithm is capable to filter out such noise and recover the original phase variance.

## 4. Experimental results

Next, we test if the optimality of our algorithm still holds when applied to experimental data. We compare the proposed ML framework with the same conventional approach described in the previous section. The optical phase tracking is performed on a digitized quadrature obtained from a down-mixed electro-optical frequency comb. The down-converted comb consists of 49 comb lines, spaced 50 MHz and centred at ca. 4.5 GHz, obtained using the same setup as in [5], for which we show the signal power spectrum in Fig. 1b.

In Fig. 4, we compare the differential phase variance extracted using the proposed ML framework with the conventional approach. As showed in [5], we expect the differential phase to follow a quadratic trend, which can also be seen on the variance per line of both methods. However, we observe a discrepancy between the two curves, similar to the one seen in Fig. 3. Our ML method reveals a clearer quadratic-dependent-line variance than the conventional method. This indicates that the proposed method is more robust to measurement noise, and that it provides a better phase estimate.

## 5. Conclusions

We have introduced, and numerically as well as experimentally investigated a novel machine learning based framework for accurate noise characterization of optical frequency combs using dual-comb spectroscopy measurements. It has been demonstrated numerically and experimentally that the proposed framework provides more accurate phase noise characterization compared to the conventional approach and provides the degree of phase correlation over the full bandwidth in a line-resolved manner. The method holds the potential to become a reference tool for frequency noise characterization that will benefit OFC based communication systems.